# The influence of microstructures and crystalline defects on the superconductivity of $MgB_2$

A. Serquis, X. Y. Liao, Y. T. Zhu, J. Y. Coulter, J. Y. Huang, J. O. Willis,

D. E. Peterson, and F. M. Mueller

Superconductivity Technology Center, MS K763, Los Alamos National Laboratory, Los Alamos, NM 87545, USA

N. O. Moreno and J. D. Thompson

Condensed Matter and Thermal Physics, MS K764, Los Alamos National Laboratory, Los Alamos, NM 87545, USA

S. S. Indrakanti and V. F. Nesterenko

Department of Mechanical and Aerospace Engineering, University of California, San Diego, La Jolla, CA 92093

This work studies the influence of microstructures and crystalline defects on the superconductivity of $MgB_2$, with the objective to improve its flux pinning. A $MgB_2$ sample pellet that was hot isostatic pressed (HIPed) was found to have significantly increased critical current density ($J_c$) at high fields than its un-HIPed counterpart. The HIPed sample had a $J_c$ of 10000 A/cm$^2$ in 50000 Oe (5 T) at 5K. This was 20 times higher than that of the un-HIPed sample, and the same as the best $J_c$ reported by other research groups. Microstructures observed in scanning and transmission electron microscopy indicate that the HIP process eliminated porosity present in the $MgB_2$ pellet resulting in an improved intergrain connectivity. Such improvement in intergrain connectivity was believed to prevent the steep $J_c$ drop with magnetic field H that occurred in the un-HIPed $MgB_2$ pellet at H > 45000 Oe (4.5 T) and T = 5 K. The HIP process was also found to disperse the MgO that existed at the grain boundaries of the un-HIPed $MgB_2$ pellet and to generate more dislocations in the HIPed the pellets. These dispersed MgO particles and dislocations improved flux pinning also at H<45000 Oe. The HIPing process was also found to lower the resistivity at room temperature.

74.70.Ad, 74.60.Ge, 74.62.Bf, 74.25.Fy

## I. INTRODUCTION

The discovery of superconductivity at 39 K in $MgB_2$ by Nagamatsu et al.[1] has attracted the attention of numerous researchers, especially those in applied superconductivity. One of the challenges is understanding the grain boundary properties of the $MgB_2$ phase: whether or not weak links are a limiting factor for intergrain critical currents, similar to the situation for polycrystalline high-$T_c$ superconductors (HTSC). Several studies indicate that a strong intergranular current network is established in $MgB_2$ material, so that the current is not limited by weak-link boundaries. For example, by means of magnetization measurements of grain agglomerates, Bugolavsky et al.[2] have shown that within these microscopic structures, intergrain and intragrain critical currents are quite comparable in value. In addition, several authors[3,4,5] reported that high density samples have high superconducting homogeneity and strong intergranular current flow as determined by magneto-optical studies. However, a rapid drop of the critical current ($J_c$) at high fields, which could be related to weak link behavior, can be seen in most studies.[6]

Several works show that the transport properties of $MgB_2$ are sample dependent. Both the reported resistivity at room temperature $\rho(297\ K)$ and the residual resistivity ratio, RRR=$\rho(297\ K)/\rho(40\ K)$ vary among different research groups by about an order of magnitude.[1,4,7] In addition, the $T_c$ and the $J_c$ and its dependence on magnetic field are also sample dependent. This sample dependency has been previously attributed to the synthesis conditions (pressure or thermal history),[8] or to the presence of Mg deficiencies or to the presence of oxygen-related defects.[9]

Several methods have been reported to obtain well-connected and dense samples, and high-pressure synthesis seems to be able to produce fully dense bulk $MgB_2$ with electrical transport properties superior to those of sintered samples.[3,10] In a previous work we have produced samples with very sharp superconducting transitions using a novel Mg vapor infiltration technique.[11] We have observed that Mg(O,B) precipitates provide good flux pinning in these samples leading to a $J_c$ of 1.5 x $10^5$ A/cm$^2$ at 5 K and 1 T,[12] which is better than that of the first works published on powders and wires. [13,14,15] In this work we have used the hot isostatic pressing (HIPing) process to further improve flux pinning of our samples. The objective of this study is to understand the influence of the microstructures and crystalline defects produced by HIPing on the superconductivity of MgB$_2$.

## II. EXPERIMENTAL

MgB$_2$ samples were synthesized using an atomic ratio of Mg:B = 1:1 at 900°C under flowing Ar. The starting materials were amorphous boron powder (-325 mesh, 99.99% Alfa Aesar) and Mg turnings (99.98% Puratronic). The boron powder was pressed to pellets (5 mm diameter x 4 mm thickness). The pellets and the Mg turnings were wrapped in Ta foil, placed in an alumina crucible inside a tube furnace under ultra-high purity flowing Ar, and heated at 900°C for two hours. We shall hereafter refer to the as-synthesized MgB$_2$ pellet as the un-HIPed sample. The un-HIPed sample was ground into powder and hot isostatically pressed (HIPed) at 200 MPa to form the HIPed sample. The HIPing was carried out in an ABB Mini-HIPer unit using a cycle cooling under pressure with a maximum temperature of 1000 °C, as described elsewhere.[10]

A SQUID magnetometer (Quantum Design) was used to measure the susceptibility of the samples over a temperature range of 5 to 45 K in an applied field of 10 Oe. Magnetization versus magnetic field (M-H) curves were measured on rectangular-shaped samples at temperatures of 5 and 30 K under magnetic fields up to 70000 Oe to determine the critical current density Jc(H).

The direct current (dc) resistivity as a function of temperature was measured through the standard 4-probe method in a computer-controlled data logger system, on samples with dimensions of ~2 x 0.5 x 0.8 mm.

The surface morphology and microstructures of the samples were characterized using a JEOL 6300FX scanning electron microscope (SEM) and a Philips CM300 transmission electron microscope (TEM) operated at 300 kV and a JEOL 3000F TEM operating at 300 kV. TEM samples were prepared by grinding the $MgB_2$ pellets mechanically to a thickness of about 50 μm and then further thinning to a thickness of electron transparency using a Gatan precision ion polishing system with $Ar^+$ accelerating voltage of 3.5 kV.

## III. RESULTS AND DISCUSSION

Figure 1(a) shows the dc magnetization (M) as a function of temperature for the un-HIPed and HIPed samples. The magnetization of the commercial Alfa Aesar $MgB_2$ powder and that of the same powder HIPed from reference [7] are also shown for comparison. It is obvious that the samples synthesized in this study (both un-HIPed and HIPed) have higher $T_c$ and sharper superconducting transitions than the as-purchased and HIPed commercial Alfa Aesar samples. It can be seen that the superconducting transition in the un-HIPed sample is slightly sharper than those of the HIPed samples. It is not clear what caused the transition broadening in the HIPed sample. However, the broadening could be related to the microstructural changes induced by the HIPing process.

As shown in Fig 1(b), if the un-HIPed sample is ground in an agate mortar resulting in particles with average size of 5-10 μm, the superconducting transition becomes broader. The same behavior is observed if the HIPed sample is ground. The transition broadening in ground samples can be explained by the penetration length λ(T) and its dependence on temperature, which follows the relationship:

$$\lambda(T) = \frac{\lambda(0)}{\sqrt{1-(T/Tc)^2}} \qquad (1)$$

where $\lambda(0)$ = 110-180 nm,[14,15,16,17] for $MgB_2$. Therefore, the penetration length can be significantly larger near $T_c$ than the particle size (leading to transition broadening) and decreases with temperature. As shown in Fig. 2 isolated particles with sizes less than 1 μm exist in the ground un-HIPed sample. These particles are smaller than the penetration length near Tc, and became superconducting at lower temperature, giving rise to the transition broadening. In other words, the transition broadening is caused by the small particle size rather than sample inhomogeneity, impurities, or weak link behavior as was suggested by Rogado et al.[18]

The resistivity, ρ(T), of the HIPed and un-HIPed samples are shown in Fig. 3. The onset of the transition is 39.4 K for the un-HIPed sample and 38.5 K for the HIPed sample, while the transition width ΔT(10% to 90%) is 0.3 K for both samples. The lower resistivity at room temperature ρ(297 K) for the HIPed sample may be due to its higher density and better inter-granular connections, as will be discussed later. The RRR=ρ(297 K)/ρ(40 K) of the un-HIPed and the HIPed samples are ~8.9 and 3.1, respectively. Lee et. al.[19] reported that single crystals of $MgB_2$ (~100 μm) with superconducting transitions around 38.1-38.3 K and ΔT = 0.2-0.3 K have an estimated resistivity at 40 K of 1 μΩcm and a RRR=5. As the phonon contribution to resistivity decreases with decreasing temperature, other defects present inside the grains (not in the grain boundaries) must have affected the resistivity near Tc. Xue et al[9] reported a correlation between RRR and the strain determined by Rietveld analysis of the x-ray diffraction data. In Fig. 4 we plot the strain values vs. the RRR of our samples,[11] together with the data of Xue et al[9]. The clear dependence of RRR upon strain confirms that the residual resistance is related to lattice distortion (strain). As we can see below, this lattice distortion may be related to the presence of higher densities of defects.

The magnetization versus magnetic field (M-H) curves of the HIPed and the un-HIPed samples is shown in Fig. 5. It can be seen from the insets that the reversible region in the HIPed sample lies above 7 T at T = 5 K and above 1.9 T at 30 K, while for the un-HIPed sample, these values are reduced to 6.2 T and 1.8 T, respectively.

Jc(H) was determined using the Bean critical state model[20] for a long parallelepiped:

$$Jc(H)[A/cm^2] = \frac{20 \times \Delta M(H)[emu/cm^3]}{(a - \frac{a^2}{3b})[cm]} \qquad (2)$$

where a and b are the lengths of the parallelepiped edges perpendicular to the magnetic field and ΔM is the width of the magnetization characteristic at the applied magnetic field *H*.

Figure 6 shows the dependence of $J_c$ on the applied magnetic field for both HIPed and un-HIPed samples. The $J_c$ at 0 T is nearly the same for both samples. However, the HIPed sample has a significantly higher $J_c$ than the un-HIPed sample in magnetic field. The difference in $J_c$ between the HIPed and un-HIPed samples increases with field. The un-HIPed sample shows a steep drop in $J_c$ at higher fields (H > 45000 Oe and T = 5 K). No such steep drop in $J_c$ is observed in the HIPed sample.

In fact, the HIPed sample has better flux pinning than samples reported earlier by other groups and is among the best reported. The excellent flux pinning in the HIPed sample was caused by the modification of microstructures and crystalline defects during the HIPing process. Therefore, it is essential to study the differences in the microstructures and defects of the HIPed and un-HIPed $MgB_2$ samples. Shown in Fig. 7(a) is the surface morphology of the un-HIPed sample. Well-developed grains of around 0.3-5 μm can be seen in the figure. However, the grains are not well connected on the surface. In contrast, the grains in the HIPed sample (Fig. 7(b)) are well compacted. The high density of the HIPed sample makes it possible to prepare a polished shiny surface of mirror quality (Fig. 7(c)).

Figure 8(a) shows a TEM bright-field image of the un-HIPed sample, which reveals poor connectivity among the $MgB_2$ grains. The white areas in the figure are pores, and the dark areas have been proven to be MgO by electron diffraction, as shown in Fig. 8(c). The MgO consists of nanometer-sized grains, as demonstrated in Fig. 8(b). The images in Figs. 7(a) and 8(a) clearly show that the $MgB_2$ grains of the un-HIPed sample are not well connected. In other words, weak links exist at the grain boundaries of this sample. Figure 9 shows the excellent connectivity between the $MgB_2$ grains of the HIPed sample. Neither pores nor MgO are seen at $MgB_2$ grain boundaries. The MgO, seen at the $MgB_2$ grain boundaries in the un-HIPed sample (Fig. 8(a)), have been broken up and dispersed inside the $MgB_2$ grains in the form of fine MgO particles. In addition, the dislocation density inside the $MgB_2$ grains is much higher in the HIPed sample (Fig. 9) than in the un-HIPed sample (Fig. 8).

In a previous work,[12] we also observed nanometer-sized coherent Mg(B,O) precipitates inside the $MgB_2$ grains in the un-HIPed samples, which are also present in the HIPed ones. However, because of the non-zone-axis imaging conditions used for Figs. 8(a) and 9, the precipitates cannot be seen in these images. Note that the fine MgO particles, with sizes around 10 to 50 nm, as well as high density of dislocations, are present only in the HIPed sample. It is well known that extended defects, such as dislocation networks and small precipitates of non-superconducting second phases, are likely to be effective flux pinning centers. In other words, the higher $J_c$ in the HIPed sample (Fig. 6) is due to effective flux pinning of finely dispersed MgO oxides particles as well as high-density dislocation networks. These finely dispersed MgO particles and high density dislocations also caused the higher strain in HIPed sample, which resulted in a low RRR value in that sample.

The steep drop in $J_c$ at higher fields (>45000 Oe) in the un-HIPed sample is attributed to the porosity and MgO oxide on the grain boundaries. Dou et al.[21] reported evidence for decoupling of the grains in sintered $MgB_2$, both through partial flux jumping and a step in the field dependence of Jc. These authors showed that initial bulk superconductor samples break down into a granular assembly beyond a certain critical value of field and temperature. As well as in our un-HIPed sample they observed a clear drop in the $\Delta M$ (which is proportional to the Jc), which can be attributed to the magnetic breakdown of the grain matrix as a result of flux penetration into the grain boundaries that may contain impurities. In the low-field region the current circulates mainly over the entire sample size (intergranular current), while in the high field region the current circulates only in the individual grains (intragranular current). It seems that for the un-HIPed sample this decoupling occurs ~ 45000 Oe at 5 K, whereas for the HIPed sample we did not observe a steep drop in the $J_c$ up to 70000 Oe.

## IV. SUMMARY

In summary, both the sample sintered at ambient pressure (un-HIPed) and the one that was HIPed present the same superconducting properties at low fields, with a very sharp transition observed by both magnetization and resistivity values. The lower RRR values can be ascribed to the presence of strain. Besides, the unHIPed sample contains discernible empty space (pores) as well as impurity phases at the grain boundaries. The weak connectivity between domains and the presence of impurities in the grain boundaries in the $MgB_2$ ambient temperature sintered sample seem to limit $J_c$ at high fields, although not as severely as in high-$T_c$ superconductors. The HIP process improves the field dependence of $J_c$ through better connectivity of the grains, the generation of dislocations, and the destruction of MgO at $MgB_2$ boundaries, which is redistributed in the form of fine particles inside the $MgB_2$ matrix. These defects can act as effective flux pinning centers. The HIPed sample also has a higher irreversibility field, which is an important parameter in potential applications.

## ACKNOWLEDGMENT

Work at Los Alamos was performed under the auspices of the US DOE Office of Energy Efficiency and Renewable Energy, as part of its Superconductivity for Electric Systems Program.

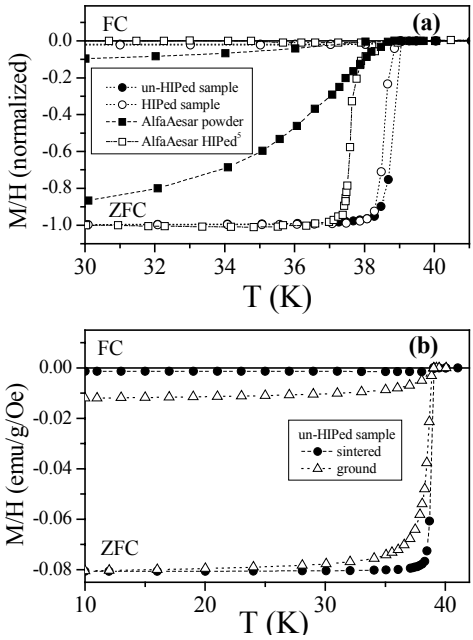

Fig. 1- (a) Magnetic susceptibility as a function of temperature of $MgB_2$ un-HIPed sample and the HIPed sample. For comparison we have added the data from reference [7] of the commercial Alfa Aesar $MgB_2$ powder and the same powder HIPed. (b) Magnetic susceptibility as a function of temperature of $MgB_2$, where the superconducting transition of sintered un-HIPed sample is compared with that of the same sample ground.

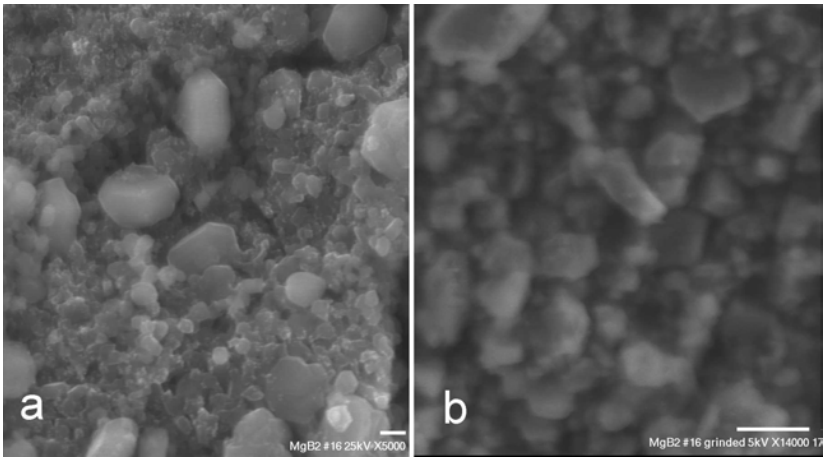

Fig. 2- SEM micrographs of (a) sintered and (b) ground un-HIPed sample.

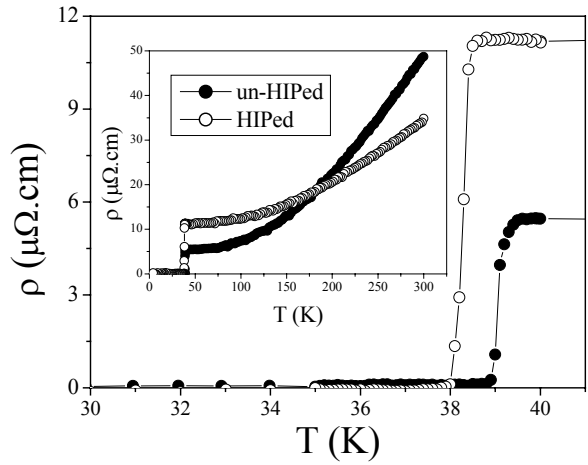

Fig. 3- Dc resistivity as a function of temperature for the un-HIPed and HIPed samples. The inset show an extended range of temperature.

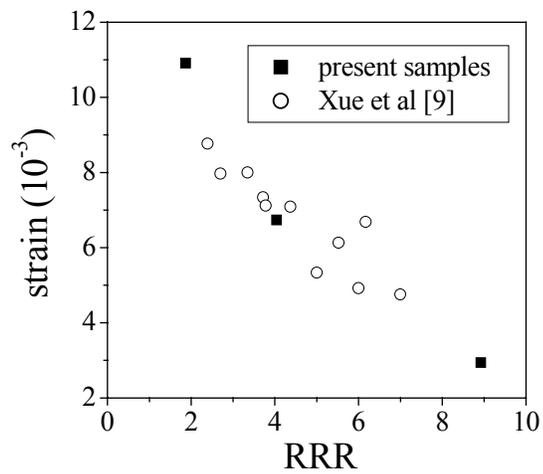

Fig. 4 - The strain vs. RRR=$\rho(297\ K)/\rho(40\ K)$ of the present samples reported in reference [11]. For comparison we show the results of Xue et al.[9]

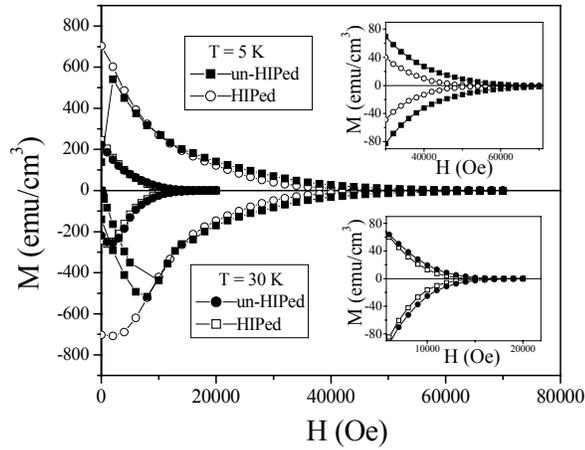

Fig. 5- Magnetization M as a function of magnetic field H at 5 and 30 K for the un-HIPed and HIPed samples. Upper inset: expanded view showing the onset of the reversible regime at 5 K. Lower inset: expanded view showing the onset of the reversible regime at 30 K.

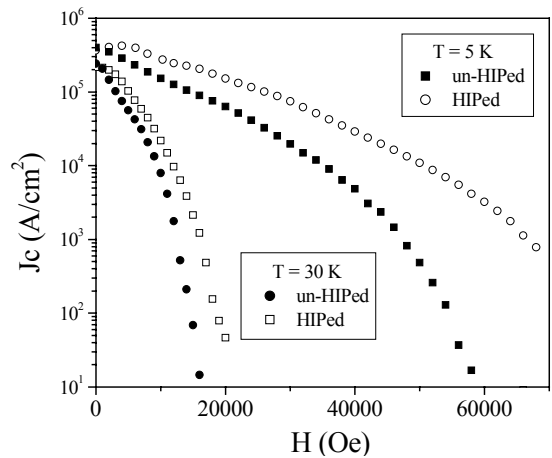

Fig. 6- Magnetization critical current density $J_c$ as a function of magnetic field H for the un-HIPed and HIPed samples at 5 and 30 K. As discussed in the text, the $J_c$ at 0 field is nearly the same for both samples, but the differences between the samples increases with field, and the drop in $J_c$ at higher fields is remarkably faster in the un-HIPed sample than in the HIPed one

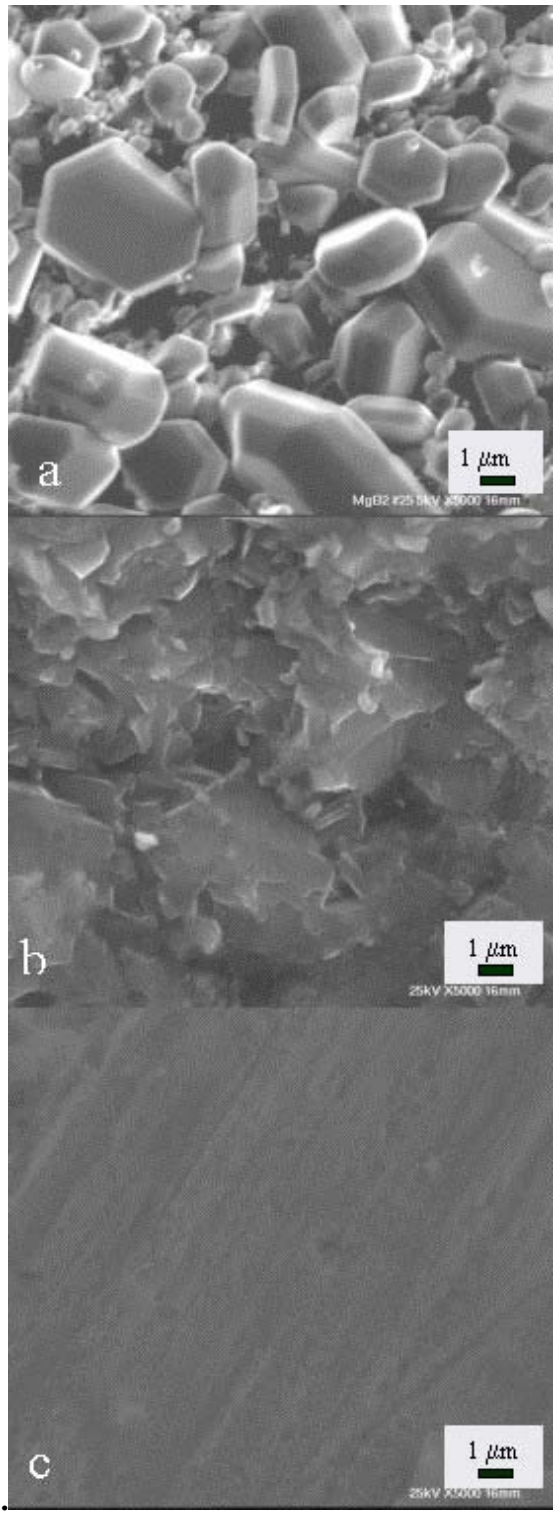

Fig. 7- SEM micrographs of MgB$_2$ samples: (a) surface of un-HIPed sample, (b) HIPed sample and (c) polished HIPed sample.

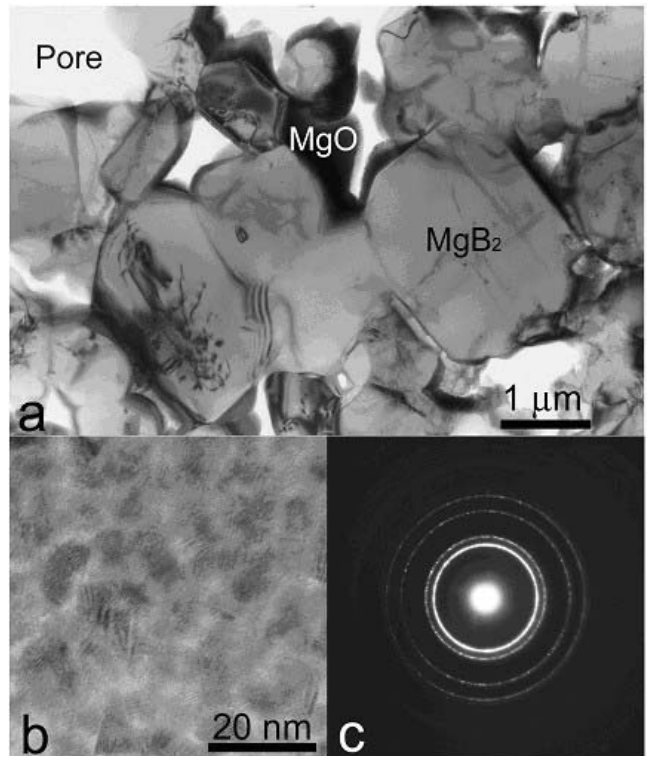

Fig. 8- (a) A bright-field image of the un-HIPed sample reveals poor connectivity among $MgB_2$ grains. Pores and MgO at the grain boundaries of $MgB_2$ are seen; (b) a magnified image of a MgO area showing the nanometer-sized grains characteristic of MgO; (c) electron diffraction pattern of MgO from one of these areas.

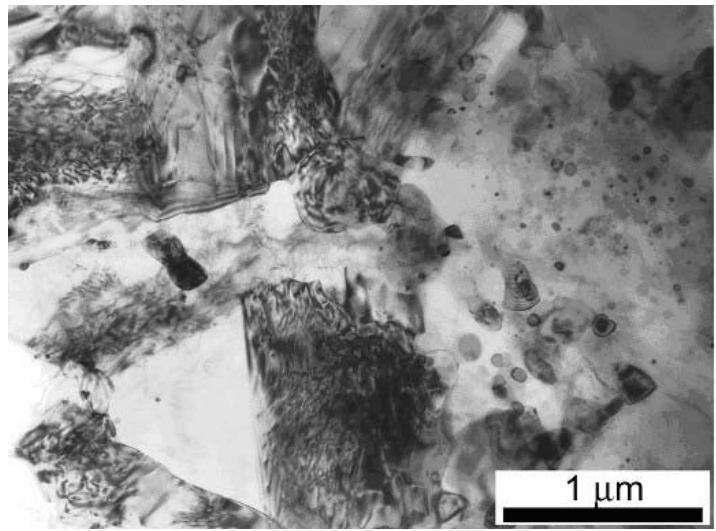

Fig. 9- A bright-field image of the HIPed sample shows that connectivity among the $MgB_2$ grains has been greatly improved. No pores are seen.